\documentclass[12pt,titlepage]{article}

\usepackage{comment}
\usepackage{color}
\usepackage{setspace}
\usepackage{graphicx}
\usepackage{epstopdf}
\usepackage{bbm}
\usepackage{rotating}
\usepackage{textcomp}

\usepackage{amsfonts}
\usepackage{amssymb}
\usepackage{amsmath}
\usepackage{graphicx}
\usepackage{dsfont}
\usepackage{epsfig}
\usepackage{ae}
\usepackage{hyperref}
\usepackage{harvard}
\usepackage{amsmath}

\newtheorem{theorem}{Theorem}
\newtheorem{lemma}{Lemma}
\newtheorem{Prop}{Proposition}
\newtheorem{assu}{Assumption}

\newenvironment{proof}[1][Proof]{\noindent\textbf{#1.} }{\ \rule{0.5em}{0.5em}}
\newtheorem{theo}{Theorem}[section]
\newtheorem{pr}{Proposition}[section]
\newtheorem{co}{Corollary}[section]

\newtheorem{exa}{Example}[section]
\newtheorem{lem}{Lemma}[section]

\def\proof{\noindent \textbf{Proof:}\, }

\def\be{\begin{equation}} 
\def\ee{\end{equation}} 
\def\beqn{\begin{eqnarray}} 
\def\eeqn{\end{eqnarray}} 
\def\beq{\begin{eqnarray*}} 
\def\eeq{\end{eqnarray*}} 
\def\ba{\begin{array}} 
\def\ea{\end{array}} 
\newcommand{\bt}{\begin{theo}}
\newcommand{\et}{\end{theo}}
\newcommand{\bl}{\begin{lem}}
\newcommand{\el}{\end{lem}}
\newcommand{\bc}{\begin{co}}
\newcommand{\ec}{\end{co}}
\newcommand{\bp}{\begin{pr}}
\newcommand{\ep}{\end{pr}}
\newcommand{\bex}{\begin{exa}}
\newcommand{\eex}{\end{exa}\vspace{-4mm}}
\newcommand{\br}{\begin{re}}
\newcommand{\er}{\end{re}\vspace{-3mm}}

\definecolor{myBlue}{rgb}{0.80,0.85,1.00}
\definecolor{myYellow}{rgb}{0,1.000,0}

\newbox\treebox
\def\tree{\global\setbox\treebox=\boxtree}

\def\endsubtree{\ettext \egroup}

\newif\iftreetext\treetextfalse
\def\boxtree{\hbox\bgroup
  \baselineskip 2.5ex
  \tabskip 0pt
  \vbox\bgroup
  \treetexttrue
  \let\par\crcr \obeylines
  \halign\bgroup##\hfil\cr}
\def\ettext{\iftreetext
  \crcr\egroup \egroup \fi}

\def\cons#1#2{\edef#2{\xmark #1#2}}
\def\car#1{\expandafter\docar#1\docar}
\def\docar\xmark#1\xmark#2\docar{#1}
\def\cdr#1{\expandafter\docdr#1\docdr#1}
\def\docdr\xmark#1\xmark#2\docdr#3{\def#3{\xmark #2}}
\def\xmark{\noexpand\xmark}
\def\nil{\xmark}

\def\settreesizes{\setbox0=\copy\treebox \global\let\treesizes\nil \setsizes}
\newdimen\treewidth
\def\setsizes{\setbox0=\hbox\bgroup
  \unhbox0\unskip
  \inittreewidth
  \sizesubtrees
  \sizelevel
  \egroup}

\def\inittreewidth{\ifx\treesizes\nil   
  \treewidth=0pt                        
\else \treewidth=\car\treesizes         
  \global\cdr\treesizes                 
  \fi}                                  

\def\sizesubtrees{\loop                 
  \setbox0=\lastbox \unskip             
  \ifhbox0 \setsizes                    
  \repeat}                              

\def\sizelevel{\ifdim\treewidth<\wd0    
  \treewidth=\wd0 \fi                   
\global\cons{\the\treewidth}\treesizes} 

\newdimen\treeheight                    
\newif\ifleaf                           
\newif\ifbotsub                         
\newif\iftopsub                         
\def\maketree{\hbox{\treewidth=\car\treesizes  
  \cdr\treesizes                        
  \makesubtreebox\unskip                
  \ifleaf \makeleaf                     
  \else \makeparent \fi}}               

{\catcode`@=11                          
\gdef\makesubtreebox{\unhbox\treebox    
  \unskip\global\setbox\treebox\lastbox 
  \ifvbox\treebox                       
    \global\leaftrue \let\next\relax    
  \else \botsubtrue                     
    \setbox0\box\voidb@x                
    \botsubtrue \let\next\makesubtree   
  \fi \next}}                           

\def\makesubtree{\setbox1\maketree      
  \unskip\global\setbox\treebox\lastbox 
  \treeheight=\ht1                      
  \advance\treeheight 2ex               
  \ifhbox\treebox \topsubfalse          
    \else \topsubtrue \fi               
  \addsubtreebox                        
  \iftopsub \global\leaffalse           
    \let\next\relax \else               
    \botsubfalse \let\next\makesubtree  
  \fi \next}                            

\def\addsubtreebox{\setbox0=\vbox{\subtreebox\unvbox0}}
\def\subtreebox{\hbox\bgroup            
  \vbox to \treeheight\bgroup           
   \ifbotsub \iftopsub \vfil            
       \hrule width 0.4pt               
     \else \treehalfrule \fi \vfil      
   \else \iftopsub \vfil \treehalfrule  
     \else \hrule width 0.4pt height \treeheight \fi\fi 
   \egroup                              
  \treectrbox{\hrule width 1em}\hskip 0.2em\treectrbox{\box1}\egroup}

\def\treectrbox#1{\vbox to \treeheight{\vfil #1\vfil}}
\def\treehalfrule{\dimen0=\treeheight   
  \divide\dimen0 2\advance\dimen0 0.2pt 
  \hrule width 0.4pt height \dimen0}    

\def\makeleaf{\box\treebox}             
\def\makeparent{\ifdim\ht\treebox>\ht0  
  \treeheight=\ht\treebox               
\else \treeheight=\ht0 \fi              
\advance\treewidth-\wd\treebox          
\advance\treewidth 1em                  
\treectrbox{\box\treebox}\hskip 0.2em   
\treectrbox{\hrule width \treewidth}\treectrbox{\box0}} 

\newcommand{\TITLE}
{Smoothness of the Value Function for Optimal Consumption Model with Consumption-Wealth Utility and Borrowing Constraint}

\begin{document}

\pagestyle{plain}

 \nocite{*}
\bibliographystyle{econometrica}

\author{Weidong Tian\\
University of North Carolina at Charlotte 
\\
\and 
Zimu Zhu
\\ University of California, Santa Barbara
 }
 \date{}
 
 \renewcommand\footnotemark{}
	\title{\Large \bf  \TITLE}
	\thanks{
Corresponding author: Zimu Zhu (Email: zimuzhu@ucsb.edu), University of California, Santa Barbara . Weidong Tian (Email:wtian1@uncc.edu), Belk College of Business, University of North Carolina at Charlotte.  We greatly thank Michael Ludkovski and Jianfeng Zhang for stimulating discussions on this paper.  We are grateful to editors and an anonymous referee for their comments.}

\maketitle
%
%
%
%
%
\renewcommand{\abstractname}{
{\LARGE Smoothness of the Value Function for Optimal Consumption Model with Consumption-Wealth Utility and Borrowing Constraint}\\[1in] Abstract}

\begin{abstract}
This paper studies an optimal consumption-investment problem for an investor whose instantaneous utility depends on both consumption and wealth, and the investor faces a general borrowing constraint that the investment amount in the risky asset does not exceed an exogenous function of the wealth. We show that the value function is second-order smooth and present the optimal consumption-investment policy in a feedback form. Moreover, when the risky investment amount is bounded above by a fixed constant, we show that under certain conditions,  the constraint is binding if and only if an endogenous threshold bounds the portfolio wealth, and we determine the endogenous wealth threshold with the smooth fit condition. Our results encompass several well-developed portfolio choice models and imply new applications. 
 \vspace{0.9cm}




\end{abstract}

\section{Introduction}
%
Since the seminal study by Merton (1971) on the portfolio choice problem of a time-separable preference for consumption, numerous studies have been on the optimal portfolio choice problem on a general utility function for consumption and other endogenous variables. For instance, Bismut (1971) introduces a dual method to describe the value function for a general utility function. Federico, Gassiat, and Gozzi (2015) demonstrate the smooth properties of the value function for a general preference for consumption and endogenous wealth. In addition to the instantaneous consumption and wealth level, several researchers have incorporated the trajectory of consumption rates within a habit formation framework. Notable works in this domain include studies by Deng, Pham, and Yu (2022); Detemple and Karatzas (2013); Englezo and Karatzas (2009); Guasoni, Huberman, and Ren (2020); and van Bilsen, Laeven, and Nijman (2020). Bank and Riedel (2001) also offer insights into the value function for a broad preference for cumulative consumption. Moreover, when wealth level or a trajectory of wealth is incorporated into utility, contributions from Bakshi and Chen (1996), Barles, Daher, and Romano (1994), Bouchard and Pham (2004), and Smith (2011) are noteworthy.\footnote{Beyond time-separable preference for consumption, understanding general utility is paramount and has been explored in various economic contexts. Examples include preferences related to labor and goods as seen in Ait-Sahalia, Parker, and Yogo (2004) and Bodie, Merton, and Samuelson (1991); variations in monetary units discussed by Miao and Xie (2013); and considerations of cash investments as per Kraft and Weiss (2019). Further readings on the optimal portfolio-choice issue within a recursive utility can be found in Schroder and Skidas (2003) and Xing (2017).}

The optimal portfolio choice problem has also been widely studied under different borrowing constraints for a standard time-separable preference. Grossman and Vila (1992) and Vila and Zariphopoulou (1997) analyzed borrowing constraints on risky investments, which are linearly contingent on the wealth process. Zariphopoulou (1994) considers the constraint on the risky investment by a generally concave, increasing function of the wealth process, focusing exclusively on consumption utility. The research exploring constraints related to consumption or the wealth process is abundant. See, for instance, Black and Perold (1992), Bardhan (1994), Dybvig (1995), El Karoui and Jeanblanc-Picque (1998), Elie and Touzi (2008), Dybvig and Liu (2010), Chen and Tian (2016), Xu and Yi (2016), and Ahn, Choi, and Lim (2019), to name a few. It is worth noting that many of these studies predominantly assume time-separable preference specifications for consumption.

Remarkably, limited studies address the optimal portfolio choice problem when considering a general utility specification paired with a broad class of trading constraints. One of the complex aspects of a constrained portfolio choice problem is the {\em smoothness} of the value function. Even more challenging is obtaining analytical representations of the constraint regions. This paper delves into an optimal consumption and investment problem, bringing two novel features to the fore: (1) The investor's instant utility is affected by consumption and an additional endogenous variable - wealth. (2) A dynamic borrowing constraint exists on the risky investment, characterized by a generally concave, increasing function of the portfolio wealth. We explore this problem within a financial market context, spanning an indefinite trading horizon and encompassing both a risk-free and a risky asset. The price of the risky asset follows a Geometric Brownian Motion. In addition, shorting is allowed, but the wealth must stay nonnegative, i.e., bankruptcy is prohibited. 
%
%


This paper offers two primary findings. Firstly, we establish (as detailed in Theorem \ref{thm:Case1-smooth}) the smoothness of the value function for the optimal consumption problem. This is done within the context of a general utility function encompassing both consumption and wealth, and operates under a general borrowing constraint on risky investments. To draw parallels, for a time-separable preference for consumption, Zariphopoulou (1994) demonstrates that the value function is the singular viscosity solution to the related HJB equation. Then,  she obtains the  smoothness of the value function by verifying that the HJB equation is uniformly elliptic. 
Strulovici and Szydlowski (2015) confirm the smoothness of the value function for a stochastic control problem with a broad utility function and a particular compact control set. Put simply, both studies underscore the smoothness of the value function by affirming that the HJB equation exhibits uniform ellipticity. In generic situations regarding both preference specification and borrowing constraint, however, ensuring the condition of uniform ellipticity presents a challenge.

%

Addressing this technical challenge, we introduce a new method for proving the value function's smoothness, as delineated in Theorem \ref{thm:Case1-smooth}. Our approach combines the uniformly elliptic techniques from Zariphopoulou (1994) and Strulovici and Szydlowski (2015) with the dual method proposed by Xu and Yi (2016). The latter employs a dual transformation strategy, converting the original HJB equation into an {\em auxiliary} HJB equation. Within this dual transformation framework, we establish the smoothness of the auxiliary HJB equation, which, in turn, implies the smoothness of the value function. 

Specifically, we first characterize the value function as the unique viscosity solution of the HJB equation, aligning with the insights of Crandall, Ishii, and Lions (1992) or Fleming and Soner (2006).  Subsequently, we dissect the nonnegative wealth domain into two segments: the unconstrained and the constrained domains. For the ``constrained domain", we ascertain the value function's smoothness using the uniform elliptic condition. In the ``unconstrained domain", leveraging the Legendre-Fenchel transformation, we demonstrate that the auxiliary HJB equation is a non-degenerate, quasilinear ODE, whereas the HJB equation might be degenerate.\footnote{As will be explained below, our approach in the unconstrained region 
is different from Federico, Gassiat, and Gozzi (2015), as the unconstrained region is unknown in our setting.} This revelation substantiates the smoothness of both the auxiliary and original HJB equations.  Finally, we resolve another technical issue verifying the value function's smoothness at the junctures where the constrained and unconstrained domains intersect. We present the optimal control policy in a feedback format.

%

Theorem \ref{thm:Case1-smooth} encompasses various findings from the extant literature. Grossman and Vila (1994) characterize the optimal portfolio growth rate with a linear constraint function on wealth in a continuous-time setting.  Zariphopoulou (1994) demonstrates the $C^2$ smooth property of the value function, focusing solely on the utility function of consumption within a general borrowing constraint. This exploration was deepened by Vila and Zariphopoulou (1997), who provided a detailed characterization of the value function exclusively for CRRA utility of consumption. When wealth is viewed as luxury goods or social status, some authors have solved the optimal portfolio choice problems and derived asset pricing implications. Notable contributions in this domain come from Carroll (2000, 2002), Bakshi and Chen (1996), Roussanov (2010), and Smith (2001), each prescribing specific utility specifications. Finally,  the instantaneous utility might depend only on the wealth in different contexts of optimal portfolio choice problems, in such as  Liu and Loewenstein (2002), Tian and Zhu (2022). Theorem \ref{thm:Case1-smooth} extends these studies by affirming the smoothness of the value function when the utility is contingent on both consumption and wealth in a general specification, and the investor faces a general borrowing or leverage constraint.

The second pivotal result of this paper, as outlined in Theorem \ref{thm: Two region}, elucidates the constraint region under specific conditions. Taking into account the smoothness property of the value function, it becomes intuitive to reformulate the HJB equation into two distinct ODEs, each corresponding to the unconstrained and constrained domains, respectively. We focus on situations where the risky investment has an upper bound defined by a positive constant, from a practical perspective. Under certain conditions detailed in Theorem \ref{thm: Two region}, we establish a sufficient condition for the existence of threshold $x^*$ such that the constrained domain is exactly $(x^*,\infty)$,  and we name it as a {\em two-regions property}. Such a characterization of the constraint region and unconstraint region is crucial to analyze the consumption and investment policy, as both the threshold parameter $x^*$ and the value functions within the constrained and unconstrained domains are concurrently determined, adhering to conditions of continuity and smooth fit.

To explain our findings, we present several utility function examples.  In the context of a consumption-only preference, Theorem \ref{thm: Two region} generalizes the characterization presented by Vila and Zariphopoulou (1997). When focusing solely on wealth preference, Theorem \ref{thm: Two region} addresses an unresolved query raised by Tian and Zhu (2022). Broadly, Theorem \ref{thm: Two region} stipulates a condition wherein the demarcation between constraint and unconstraint regions is solely determined by a singular portfolio wealth threshold, $x^*$.

The rest of the paper is organized as follows. In Section \ref{sec: Model}, we formulate a continuous-time optimal consumption problem under a general consumption-wealth preference with a borrowing constraint.  In Section \ref{sec: smoothness}, we show that the value function is second-order smooth, and the optimal consumption-investment policy are derived in a feedback form. In Section \ref{sec: two region}, we show that under certain conditions the constrained domain is $(x^*,\infty)$ .  Finally, Section \ref{sec: applications}, we present some examples to illustrate our major results.  We conclude the paper in Section \ref{sec:conclusion}. Technical proofs are given in Appendix.
\section{Model Setup}\label{sec: Model}

There are two assets in a continuous-time economy with time $t \in [0, \infty)$. Let  $(\Omega, ({\cal F}_t)_{t\geq 0}, P)$  be a filtered probability space in which the information 
flow in the economy is generated by a standard Brownian motion $(B_t)_{t\geq 0}$. The risk-free asset (``bond") grows at a continuously compounded, constant $r$. 
The other asset (``the stock index") is a risky asset, and its price process $S_t$ follows
\be
dS_t = S_t (\mu dt + \sigma dB_t),
\ee
where $\mu$ and $\sigma$ are the expected return and the volatility of the stock return.  We assume $\mu > r>0$ and $\sigma>0$.

%

We denote ${\cal A}(x)$ as the set of feasible consumption-investment strategies $(c_t, \pi_t)_{t\geq 0}$ with $X_0 = x$ such that:
 
  i) There is a strong solution, $X_t = X_{t}^{(\pi, c)}$,  of the stochastic differential equation
\be
\label{eq:wealth}
dX_t = \pi_t [(\mu-r)dt + \sigma dB_t] + rX_t dt  - c_t dt,  X_0 \geq 0.
\ee
 and $X_t \ge 0$ almost surely  for all $ t\geq 0$. 

ii) $c_t \ge 0$ almost surely  for all $ t\geq 0$,  $\int_{0}^{t} c_s ds < \infty$ almost surely  for all $t\geq 0$ and $(c_t)_{t\geq 0}$ is ${\cal F}_t-$ progressively measurable. 

iii) $(\pi_t)_{t\geq 0}$ is ${\cal F}_t-$progressively measurable., $\int_{0}^{t} \pi_s^2 ds  < \infty$  almost surely  for all $ t\geq 0$.  Moreover,  $\pi_t \leq  g(X_t)$ almost surely  for all $ t\geq 0$ ,  where the investor's borrowing constraint $g(\cdot)$ is an increasing, concave,  Lipschitz continuous and twice differentiable function on $[0,\infty)$ to $[0,\infty)$.  Furthermore,  we assume there exists $L>0$ such that
 \be
 \label{eq: lowerbound for g}
 g(x)\geq L,\forall x>0.
 \ee

The investor's expected utility is given by 
\beq
 \ E \left[  \int_{0}^{\infty} e^{-\delta t} f(c_t, X_t)dt \right].
\eeq

with discount factor $\delta >0$. We make the following conditions on $f(c,x)$:

\begin{assu} \label{Assum: utility function}
\begin{enumerate}
\item $f(c,x)$ is a $C^{2}$ smooth function on $[0, \infty) \times [0, \infty) \rightarrow [0, \infty)$ with $f(0,0)=0$.
\item ${\partial f\over \partial c}(c,x)> 0,  {\partial f\over \partial x}(c,x) > 0$ for $c>0,x>0$.  
We assume $f$ satisfies the Inada's condition:  
\beq
\lim_{c\rightarrow 0, x\rightarrow 0}{\partial f\over \partial c}(c,x)=\lim_{c\rightarrow 0, x\rightarrow 0}{\partial f\over \partial x}(c,x)=\infty
\eeq
We also assume for all $x>0$,  $\lim\limits_{c\rightarrow +\infty}{\partial f\over \partial c}(c,x)=0$.
\item The Hessian matrix of the function $f(c,x)$ is negative definite. That is, ${\partial ^2 f \over \partial c^2}(c,x) < 0$ and ${\partial ^2 f \over \partial c^2}(c,x) {\partial ^2 f \over \partial x^2}(c,x) - [{\partial ^2 f \over \partial c \partial x}(c,x)]^2 > 0$. 
\item $f(c,x) \le M(1 + c^{\gamma} + x^{\gamma})$ for some positive number $M$ and $0 < \gamma < 1$.
\end{enumerate}
\end{assu}
By Assumption \ref{Assum: utility function},  it is clear that the function $f(c,x)$ is a concave function jointly with $c$ and $x$.

The optimal portfolio choice problem is to find the optimal trading strategy $(\pi_t)$ and the consumption rule $(c_t)$ such that
\be
\label{eq:V }
{V}(x):= \sup_{(c_t, \pi_t) \in {\cal A}(x)} \mathbb{E}\left[ \int_{0}^{\infty} e^{-\delta t}f(c_t, X_t)dt  \right].
\ee

To guarantee the value function $V(x)$ is well defined,  throughout this paper,  we assume that
\beq
\delta>r\gamma+{{\gamma}(\mu-r)^2\over 2\sigma^2(1-\gamma)}.
\eeq
The above condition ensures $V(x)$ is finite when $g(x)=+\infty$,  see Karatzas,  Lehoczky, Sethi and Shreve (1986).

For later use,  we define the Legendre-Fenchel transformation of the function $f(c,x)$:
\be
p(x, \zeta) = \max_{c \ge 0} \left\{ f(c,x) - c \zeta \right\},  x>0, \zeta>0.
\ee

By Assumption \ref{Assum: utility function},  there exists function $I(x,\zeta)>0$ such that 
\be
\label{eq:I}
{\partial f\over \partial c}(I(x,\zeta),x)=\zeta,  x>0, \zeta>0.
\ee
and $p(x, \zeta)=f(I(x,\zeta),x) - I(x,\zeta)\zeta $.

%
%
%
%
  
 \section{Smoothness property of the value function}
 \label{sec: smoothness}

 The following theorem is the first main result of this paper.
 
 \begin{theorem}
\label{thm:Case1-smooth}
Under Assumption \ref{Assum: utility function},  the value function $V(x)$ is the unique $C^2((0,\infty))\cup C([0,\infty))$ solution of:
\begin{equation}
\label{eq:Case1-HJB}
\delta {V}(x) = \max_{ \pi \le g(x)} \left[ (\mu-r) \pi {V}'(x) + \frac{1}{2} \sigma^2 \pi^2 {V}''(x)  \right]  \\
 + \max_{c \ge 0} \{f(c,x) - c {V}'(x)\}+rxV'(x),   (x > 0)
\end{equation}
in the class of concave functions with $V(0)=0$.  
The optimal feedback controls are given by
\beq
c^*(x)=I(x,V'(x)),  \quad \pi^*(x)=\min(-{(\mu-r)V'(x)\over \sigma^2 V''(x)},g(x)), \quad x>0. 
\eeq
where the function $I$ is defined in (\ref{eq:I}).
\end{theorem}

We briefly explain its idea, and  defer the proof of this result in the Appendix.  In Theorem \ref{thm:Case1-smooth},  we first show that $V(x)$ is the viscosity solution of equation (\ref{eq:Case1-HJB}). It is a widely acknowledged fact that $V(x)$ stands as the unique viscosity solution of the HJB equation (\ref{eq:Case1-HJB}), fundamentally deriving from Zariphopoulou (1994). In contrast with classical studies of the optimal portfolio choice problem without constraints, such as those by Federico, Gassiat, and Gozzi (2015), our approach examines the HJB equation (\ref{eq:Case1-HJB}) in two distinct regions below, demonstrating that $V(x)$ is smooth across all areas.

Precisely, we  define the unconstrained domain ${\cal U}$ such that  
 $V(x)$ is the viscosity solution of the following equation in ${\cal U}$:
\begin{equation}
\label{eq:Case1- ODE no constraint}
\delta {V}(x) = -{(\mu-r)^2\over 2\sigma^2}{(V')^2(x) \over V''(x)}+ p(x, V'(x))+ rx V'(x). 
\end{equation}
Here, $p(x, \zeta)$ is the Legendre-Fenchel transformation of the preference function $f(c,x)$.\footnote{Our definition of the unconstrained region is slightly different from standard literature such as Vila and Zariphopoulou (1997) in which $-\frac{\mu-r}{\sigma^2} \left( \frac{V'(x)}{V''(x)} \right) \le g(x)$, and we will explain its reason shortly.} Roughly speaking, 
$-\frac{\mu-r}{\sigma^2} \left( \frac{V'(x)}{V''(x)} \right) < g(x)$ in ${\cal U}$. 

Similarly,  we define the constrained domain ${\cal B}$ such that  
 $V(x)$ is the viscosity solution of the following equation in ${\cal B}$:
\begin{equation}
\label{eq:Case1-ODE with constraint}
\delta {V}(x) =   (\mu-r) g(x) V'(x) + \frac{1}{2} \sigma^2 g^2(x) {V}''(x) +p(x, V'(x))+ rx V'(x).    
\end{equation}
It informally means that $-\frac{\mu-r}{\sigma^2} \left( \frac{V'(x)}{V''(x)} \right) > g(x)$ in ${\cal B}$. 

We show that $V(x)$ is smooth in this setting and we establish it with several steps.
 {\em First}, in the constrained region, by the assumption on the function $g(\cdot)$,  there exists unique smooth solution of the equation (\ref{eq:Case1-ODE with constraint}) due to uniformly elliptic condition. {\em Second}, in the unconstrained domain,  a crucial step is to show that $V(x)$ is $C^1$ smooth.  In this step, we make use of the viscosity solution characterization. {\em Third}, we next show that $V(x)$ is $C^2$ smooth in the unconstrained region by a dual approach. {\em Fourth}, we show that the value function is $C^1$ smooth at the  connection points in $cl({\cal U}) \bigcap cl({\cal B})$. This $C^1$ property at these points follows from the $C^2$ smooth property in each (open) region: unconstrained and constrained region, and the viscosity solution characterization of the value function. {\em Fifth},  we show the $C^2$ smooth property at $cl({\cal U}) \bigcap cl({\cal B})$. {\em Finally},  a standard verification argument then concludes the proof. 


The  $C^2$ smoothness in the unconstrained domain bears relevance to the optimal consumption problem without constraint.  Federico, Gassiat, and Gozzi (2015) demonstrate the smooth properties of the value function by introducing a dual control problem. They first prove the smoothness of the value function in the dual control problem, which in turn implies the smoothness of the value function in the original control problem. However, our approach diverges from Federico, Gassiat, and Gozzi (2015) due to the unknown nature of the unconstrained region in our setting. We first establish that the value function is $C^1$ smooth, and subsequently employ a dual approach to affirm that the value function also exhibits $C^2$ smoothness. 

A general $C^2$ smooth property of the value function for $f(c,\pi, x)$ is proved in Strulovici and Szydlowski (2015), when $(c, \pi, x)$ belongs to a compact set and other technical conditions required. The general theorem of Strulovici and Szydlowski (2015) can be applied to the case that $\pi \in [ a, b]$, for $0 < a < b$.  However,  smoothness of the value function for a constraint that $g(x) = L$ remains open as $\pi$ can take any small positive real number.  In Tian and Zhu (2022), for $f(c,x) = \frac{x^{1-R}}{1-R}$, the $C^2$ smooth property of the value function is reduced to a unique solution of a non-linear equation of one-argument. For  general utility $f(c,x)$, the wealth $x$ is interpreted as luxury goods, and the optimal portfolio choice problem has been studied in Carroll (2000, 2002), Ait-Sahalia, Parker, and Yogo (2004), Watcher and Yogo (2010).  Moreover, this kind of utility is often used to model the social status of the wealthy in Bakshi and Chen (1996), Roussanov (2010), Smith (2001). But the smoothness property of the value function under borrowing constraint is not studied yet in previous studies. Theorem \ref{thm:Case1-smooth} guarantees the smoothness property of the value function for those consumption-wealth preference and a general borrowing constraint.

By Theorem \ref{thm:Case1-smooth}, the value function $V$ is $C^2$ continuous. Therefore, we can rigorously study the constrained and unconstrained region and the consumption-investment strategy. A standard way to transform the non-linear ODE (\ref{eq:Case1- ODE no constraint}) into a standard linear ODE by introducing a new variable $y = V'(x)$. By the concavity of $V(\cdot)$,  we write $x = H(y)$ for some function $H(\cdot)$.  Denote $J(y)=V(H(y))$ , then equation (\ref{eq:Case1- ODE no constraint})  in the unconstrained domain reduces to
\beq
\delta J(y) = - \frac{(\mu-r)^2}{2\sigma^2} y^2 H'(y) + p(H(y), y) + r yH(y), 
\eeq
with $J'(y) = y H'(y)$.  Given a solution of $H(\cdot)$, $\pi^* = -\frac{(\mu-r)}{\sigma^2}yH'(y) $ in terms of the variable $y$. Moreover, the optimal consumption rate $c^*$, as a function of the variable $y$,  is given by $I(H(y),y)$.

We next move to the optimal investment strategy. The optimal portion of investment in the risky asset $\overline{\pi}^*(x)$ is given by
\beq
\overline{\pi}^*(x):={\pi^*(x)\over x}
\eeq

On one hand, in the unconstrained region ${\cal U}$, the optimal proportion of wealth invested in the risky asset is given by
\beq
\overline{\pi}^*(x) = -\frac{\mu-r}{\sigma^2 x} \frac{V'(x)}{V''(x)}.
\eeq
On the other hand, in the constrained region ${\cal B}$,  the optimal proportion of wealth invested in the risky asset is 
   \beq
 \overline{\pi}^*(x) = \frac{g(x)}{x}.
  \eeq
  
  Under assumption on $g(x)$, $ \lim_{x \downarrow 0}\overline{\pi}^*(x)  = \lim_{x \downarrow 0} \frac{g(x)}{x} =+\infty$ in ${\cal B}$. The next result shows a general {\em decreasing} property of the proportion of investment in the risky asset. 
  
  \begin{lemma}
  \label{lem:proportion}
  In the constrained region, the proportion $\overline{\pi}^*(x)$ decreases with respect to the wealth.
  \end{lemma}

The following lemma shows that the unconstrained region contains an open interval $(0, x)$ for a positive number $x$; hence, the constrained region does not include $(0, a)$ for any small number $a$. 
  
  \begin{lemma}
  \label{lem:unconstrained}
  Under Assumption \ref{Assum: utility function}, there exists a positive number $x^*$ such that $(0, x^*) \subseteq {\cal U}$.
  \end{lemma}
    
  Given Lemma \ref{lem:unconstrained}, it is intended to conjecture that the constrained region is an open interval $(0, x^*)$, while the unconstrained region is $(x^*, \infty)$, for a certain positive real number $x^*$. In the next section, we will characterize the constrained and unconstrained regions accordingly.  
  
    \section{Characterization of the Constrained Region}\label{sec: two region}

In this section, we delineate the constrained region in the context of the borrowing constraint. From a practical perspective, risky investments are frequently bounded. Consequently, we focus on $g(x) \equiv L$, and the following theorem precisely characterizes the constrained and unconstrained regions, illustrating the two-regions property under specified conditions.
  
  \begin{theorem}
  \label{thm: Two region}
  Let $g(x)\equiv L$ and 
  \begin{eqnarray}
  \label{eq:def of m(x)}
  m(x)& \equiv &-(\mu-r){\partial p \over \partial x}(x,V'(x))-\sigma^2L[{\partial ^2 p \over \partial x^2}(x,V'(x))+2{\partial^2 p \over \partial x \partial \zeta}(x,V'(x))]\nonumber\\
  &&-{1\over \sigma^2L}{\partial ^2 p \over \partial \zeta^2}(x,V'(x))(\mu-r)^2(V'(x))^2+r(\mu-r)V'(x). 
  \end{eqnarray}
 Under Assumption \ref{Assum: utility function}, and if the function $m(\cdot)$ satisfies one of the following condition:
  \begin{enumerate}
  \item   $m(x)$ does not change sign on $(0,\infty)$;
   \item $m(x)$ changes the sign once and only once from positive to negative on $(0,\infty)$.
   \end{enumerate}
   then there exists a real number $x^*> 0$ such that ${\cal U}=(0,x^*)$ and ${\cal B}=(x^*,\infty)$.
  \end{theorem}
   
  To establish Theorem \ref{thm: Two region}, let
  \be
  Y(x) = (\mu-r) L V'(x) + \sigma^2 L V''(x),
  \ee
then, $Y(x) > 0$ in the unconstrained region, and $Y(x) < 0$ in the constrained region. Then, utilizing the equation on the value function in each unconstrained and constrained region, we define  elliptic operators ${\cal L}^{\cal U}$ and ${\cal L}^{\cal B}$ , respectively. These operators satisfy ${\cal L}^{\cal U}[Y] = 0$ in ${\cal U}$ and ${\cal L}^{\cal B}[Y] = 0$ in ${\cal B}$.  Subsequently, through a dedicated application of the comparison principle for the elliptic operator, and considering that the function $m(\cdot)$ changes sign at most once, we can demonstrate that ${\cal U} = (0, x^*)$, thereby refining Lemma \ref{lem:unconstrained}. 

 By Theorem \ref{thm: Two region}, an explicit solution of the optimal portfolio choice problem is reduced to find the endogenous number $x^*$, which separates the constrained and unconstrained region, thereby exhibiting a two-region property. Moreover, the critical number $x^*$  is determined by the smooth-fit condition in ${\cal U}$ and ${\cal B}$.
 Combining Theorem \ref{thm:Case1-smooth} and Theorem \ref{thm: Two region}, we obtain the following result.  
 
   \begin{Prop}
   \label{prop: two ODEs}
   Let $g(x)\equiv L$ and 
  \begin{eqnarray}
   m(x)&=&-(\mu-r){\partial p \over \partial x}(x,V'(x))-\sigma^2L[{\partial ^2 p \over \partial x^2}(x,V'(x))+2{\partial^2 p \over \partial x \partial \zeta}(x,V'(x))]\nonumber\\
  &&-{1\over \sigma^2L}{\partial ^2 p \over \partial \zeta^2}(x,V'(x))(\mu-r)^2(V'(x))^2+r(\mu-r)V'(x).  \nonumber
  \end{eqnarray}
 Under Assumption \ref{Assum: utility function}, and if the function $m(\cdot)$ satisfies one of the following condition:
  \begin{enumerate}
  \item   $m(x)$ does not change sign on $(0,\infty)$;
   \item $m(x)$ changes the sign once and only once from positive to negative on $(0,\infty)$.
   \end{enumerate}
   then there exists a real number $x^*> 0$ such that 
   
   a) ${\cal U}=(0,x^*)$ and $V(x)$ is the $C^2(0,x^*)$ solution of the following ODE
   \begin{equation}
\delta {V}(x) = -{(\mu-r)^2\over 2\sigma^2}{(V')^2(x) \over V''(x)}+ p(x, V'(x))+ rx V'(x). \nonumber
\end{equation}
   
    b) ${\cal B}=(x^*,\infty)$ and $V(x)$ is the $C^2(x^*,\infty)$ solution of the following ODE
   \begin{equation}
\delta {V}(x) =   (\mu-r) g(x) V'(x) + \frac{1}{2} \sigma^2 g^2(x) {V}''(x) +p(x, V'(x))+ rx V'(x).     \nonumber
\end{equation}

    with $V(0)=0$, $V(x^*-)=V(x^*+)$, $V'(x^*+)=V'(x^*-)$ and $V''(x^*+)=V''(x^*-)$.
   \end{Prop}
   
\section{Examples and Implications} \label{sec: applications}

In this section, we illustrate the main results with several examples. For simplicity (or view the risk-free asset as a numeraire), we assume $r=0$ throughout Section \ref{sec: applications}.

\subsection{Time-separable preference}
We start with a time-separable preference, $f(c,x) = u(c)$.

%
\begin{Prop}
\label{ex: utility only c}
Assume $g(x)\equiv L$, $f(c,x) = u(c)$ for any increasing and concave function $u(\cdot)$ satisfying Assumption \ref{Assum: utility function}. Then, the value function is second-order smooth and the two-regions property holds.
\end{Prop}


The first part of Proposition \ref{ex: utility only c} is known by Zariphopoulou (1994). The second part is entirely new and robust for a general time-separable preference specification $u(c)$.
 If $u(c)=\frac{c^{1-R}}{1-R}, 0<R<1$ and $g(x) = kx + L, k \ge 0, L > 0$, Vila and Zariphopoulou (1997) show the two-regions property for $k > 0$, and under the assumption that
\beq
\delta + \frac{(\mu-r)^2}{2 \sigma^2} > r + \frac{k (\mu-r)}{2}.
\eeq
Tian and Zhu (2022) derive the two-regions property for $k = 0$ under a weaker condition. Proposition \ref{ex: utility only c} presents the two-regions property for a general time-separable preference specification  $u(c)$.

\subsection{Preference for the endogenous wealth level}




We next consider the preference for the wealth level, that is, $f(c,x) = u(x)$.

\begin{Prop}
\label{co:2}
Let $g(x)\equiv L$, $f(c,x) = u(x)$ for any increasing and concave function $u$ satisfies Assumption \ref{Assum: utility function} . If either (1) $ \mu u'(x) + \sigma^2Lu''(x)$ does not change sign on $(0,\infty)$, or 
(2) $ \mu u'(x) + \sigma^2Lu''(x)$ changes the sign once and only once from negative to positive on $(0,\infty)$, then the value function is second-order smooth and the two-regions property prevails.
\end{Prop}


Proposition \ref{co:2} characterizes the constrained region explicitly by a positive number $x^*$ for a general utility function $u(\cdot)$. As an illustrative example, let $u(x)=\frac{x^{1-R}}{1-R}, 0<R<1$.
In the absence of the leverage constraint, it has been studied in Liu and Loewenstein (2002) for the optimal portfolio choice problem. For a finite positive number $L$, it is easy to see that
\beq
m(x)=-x^{-R-1}[\mu x-\sigma^2RL],
\eeq
which exactly changes the sign once from positive to negative on $(0,\infty)$. Therefore, the two-regions property holds.\footnote{Tian and Zhu (2022) demonstrates this two-regions property under certain technical condition. We remove this condition here as a special case of Proposition \ref{co:2}.} Moreover, by combing with Tian and Zhu (2022), we obtain an explicit expression of the consumption and investment strategy. Proposition \ref{co:2} characterizes the class of the function $u(x)$ for which the two-region property prevails.

%





Next, we consider some specifications of $f(c,x)$ with both the consumption and wealth involved.

\subsection{Additive specification}

Th following result is about an additive specification of $f(c, x)$ on both the consumption rate and the wealth,

\begin{Prop}
\label{prop:additive}
 Assume $f(c,x) = \alpha u(c)+\beta v(x)$ where $u(\cdot)$ and $v(\cdot)$ are increasing and concave functions satisfy Assumption \ref{Assum: utility function}.  Let $g(x) \equiv L$, and 
\beq
m(x)=-\beta[\mu v'(x)+\sigma^2Lv''(x)]+{1\over \alpha\sigma^2L}K'({V'(x)\over \alpha})\mu^2(V'(x))^2.
\eeq
where $K(\zeta)=(u')^{-1}(\zeta)$.  If $m(x)$ satisfies the conditions in Theorem \ref{thm: Two region}, then the two-regions property holds.
\end{Prop}

In this additive specification, investigating the two-region property is transformed into the study of the function $m(x)$ and thus the value function $V(x)$ in an explicit way. For example, consider the function 
\beq
f(c,x) = \frac{\alpha c^{1-R} + \beta x^{1-R}}{1-R},
\eeq
 i.e., $u(c)={c^{1-R}\over 1-R}, v(x)={x^{1-R}\over 1-R}$  for $0<R<1$. In such a case, the value function is smooth for the consumption-wealth preference and a constant borrowing constraint.  Furthermore, through computation, we have
\beq
m(x)=-x^{-R-1}[\mu x-\sigma^2RL]-\alpha^{{1\over R}}{\mu^2\over \sigma^2 LR}(V')^{1-{1\over R}}.
\eeq
If $m(x)$ satisfies the condition in Theorem \ref{thm: Two region}, then the two-regions property is upheld. This example is explored in detail by Tian and Zhu (2022). A numerical result of Proposition \ref{prop: two ODEs} can also be found in Tian and Zhu (2022).




\subsection{Multiplicative specification}
Similarly, we can consider a multiplicative specification of the preference studied in Bakshi and Chen (1996). We consider the function
\beq
f(c,x) =  \frac{ [c^{a} x^{b}]^{1-R}}{1-R},
\eeq
 where $a> 0,b > 0$ and $a + b < 1,0<R<1$. When $g(x)=\infty$, an explicit expression of the value function is given in Bakshi and Chen (1996).  For a general borrowing constraint, Theorem \ref{thm:Case1-smooth} states that the value function for such a preference. Moreover, by computation, 
\beq
p(x,\zeta)=a^{1\over 1-a(1-R)}[{1\over a(1-R)}-1]\zeta^{a(1-R)\over a(1-R)-1}x^{b(1-R)\over 1-a(1-R)},
\eeq
and
\begin{eqnarray}
m(x)&=&a^{1\over 1-a(1-R)}[{1\over a(1-R)}-1](V'(x))^{a(1-R)\over a(1-R)-1}x^{{b(1-R)\over 1-a(1-R)}-2}\cdot n(x), \nonumber
\end{eqnarray}
where
\begin{eqnarray}
    n(x) & \equiv &-\mu{b(1-R)\over 1-a(1-R)}x-{\mu^2\over \sigma^2L}{a(1-R)\over a(1-R)-1}({a(1-R)\over a(1-R)-1}-1)x^2\nonumber\\
&&-\sigma^2L\big[{b(1-R)\over 1-a(1-R)}({b(1-R)\over 1-a(1-R)}-1)+2{b(1-R)\over 1-a(1-R)}{a(1-R)\over a(1-R)-1}(V'(x))^{-1}x\big]. \nonumber\\\nonumber
\end{eqnarray}
Clearly, $m(x)$ has the same sign with $n(x)$ by the assumption. If $m(x)$, or equivalently for $n(x)$, satisfies the condition in Theorem \ref{thm: Two region}, then the two-regions property holds for a constant borrowing constraint.

%
%

%

\section{Conclusion} \label{sec:conclusion}
This paper elucidates the general and robust smoothness property of the value function associated with an optimal consumption problem. In this context, the investor derives utility from instantaneous consumption-wealth and confronts a general dynamic borrowing constraint. Our proof leverages a combination of the uniformly elliptic approach and the dual approach to substantiate this robust second-order smooth property. Moreover, we establish that under certain conditions, the constrained domain is given by $(x^*,\infty)$, where $x^*$ denotes a specific threshold that is determined by a smooth fit condition. We further illustrate our results with several examples. 


\newpage

\renewcommand {\theequation}{A-\arabic{equation}} \setcounter
{equation}{0}
\section*{Appendix: Proofs}

 We start with  viscosity solution characterization of the value function. Briefly speaking, $V(x)$ is a viscosity subsolution of an elliptic second-order equation $F(x, u, u_x, u_{xx}) = 0$ if for any smooth function $\psi$ and a maximum point $x_0$ of $V - \psi$, the inequality 
 \beq
 F(x_0, V(x_0), \psi_{x}(x_0), \psi_{xx}(x_0)) \le 0
 \eeq
holds.  Similarly, $V$ is a viscosity supersolution if for any smooth function $\psi$ and a minimum point $x_0$ of $V - \psi$, the inequality 
\beq
F(x_0, V(x_0), \psi_{x}(x_0), \psi_{xx}(x_0)) \ge 0
\eeq
holds.  A viscosity solution is both a viscosity subsolution and supersolution. We refer to Fleming and Soner  (2006) for the theory of viscosity solution. 
 
\begin{Prop}
\label{prop:Case1-vis}
The value function $V(x)$ is continuous, strictly increasing and strictly concave.  Moreover,  $V(x)$ is the unique viscosity solution of
\begin{equation}
\label{eq:Case1-vis}
\delta {V}(x) = \max_{  \pi \le g(x)} \left[ (\mu-r) \pi {V}'(x) + \frac{1}{2} \sigma^2 \pi^2 {V}''(x)  \right]  \\
 + \max_{c \ge 0} \{f(c,x) - c {V}'(x)\}+ rx V'(x),   (x > 0)
\end{equation}
in the class of concave functions with $V(0)=0$. 
\end{Prop}
\proof The concave property of the value function follows from the concavity property of $f(c,x)$ and concavity property of  $g(x)$.  The part $V(0)=0$ is standard.  For the viscosity solution characterization of the value function $V(x)$, this is a deep theorem of Zariphopoulou (1994), in which we replace $u(c)$ by  $f(c,x)$ throughout the entire proof.  \hfill$\Box$

The challenge of proving Theorem \ref{thm:Case1-smooth} is to show that the viscosity concave solution is $C^2$. We need several preliminary results to demonstrate the smooth properties. 

 The first lemma is related to the derivative of a concave function.  Since we cannot find the proof of this known result in available reference, we present its proof here.
\begin{lem}
\label{lem: c1-property}
Assume $f(x)$ is a concave function on $[a,b]$ for some real numbers $a$ and $b$.  If $f(x)$ is differentiable on $(a,b)$,  then $f(x)$ is $C^1$ on $(a,b)$.
\end{lem}

\proof   Since $f(x)$ is differentiable and concave on $(a,b)$,  $f'(x)$ is a decreasing function on $(a,b)$.  Fix $x_0 \in (a,b)$,  for all $\epsilon >0$, by Darboux theorem, there exists $x_1 \in (x_0,{x_0+b \over2})$ such that 
\beq
f'(x_1)=\max(f'(x_0)-\epsilon, f'({x_0+b\over 2}))
\eeq
therefore, by $f'(x)$ is decreasing,  we have for all $x\in (x_0,x_1)$
\be
\label{eq: rightcont}
0\leq f'(x_0)-f'(x)\leq \epsilon
\ee
Similarly,  for same $\epsilon$, we could find $x_2\in ({a+x_0\over2},x_0)$ such that when $x\in (x_2,x_0)$, we have
\be
\label{eq: leftcont}
0\leq f'(x)-f'(x_0)\leq \epsilon
\ee
Combining (\ref{eq: rightcont}) and  (\ref{eq: leftcont}), we have $f$ is $C^1$ at $x_0$.  Since $x_0$ is arbitrary,  we have $f$ is $C^1$ on $(a,b)$.        $\hfill \Box$

The next lemma is given in Tian and Zhu (2022). The proof below belongs to Prof. Jianfeng Zhang.  
\begin{lem}
\label{lem:smooth-fit}
Let $F: (0, \infty)  \times \mathbb{R} \times \mathbb{R} \times \mathbb{R} \rightarrow \mathbb{R}$ be a continuous and elliptic operator, that is,
 $F(x,r,p, X) \leq F(x,r,p, Y), \forall X \ge Y$. 
Assume $V(x)$ is a continuous and concave viscosity solution of a second-order (HJB) equation $F(x, u, u_x, u_{xx}) = 0$ and the region of $x$ is ${\cal D} = (0, \infty)$.  For $x^*\in (0,\infty)$ if there exist two numbers  $0<x_1<x^*,x^*<x_2<\infty$, such that $V(x)$ is smooth in $(x_1, x^*)$ and $(x^*, x_2)$, then $V(x)$ must satisfies the smooth-fit condition at $x^*$, that is, $V'(x^*-) = V'(x^*+)$.
\end{lem}

\proof Since $V(x)$ is smooth in both $(x_1,x^*)$ and $(x^*,x_2)$,  by the concavity of $V$,  $V'(x^{*}-)$ and $ V'(x^{*}+)$ exist.  Without lost of generality, we assume that $V'(x^{*}-) < 0 < V'(x^{*}+)$ and derive a contradiction. Since there is no available test function, the subsolution holds automatically.  We next check the supersolution. Let the test function in the form of 
\beq
\psi(x) \equiv V(x^*) + \frac{1}{2} \left[V'(x^{*}-) + V'(x^{*}+)  \right](x-x^*) + \alpha(x-x^*)^2
\eeq
We claim that $\alpha$ can take any real value: To make $\psi(x)$ the valid test function, we need to guarantee that $ \psi(x)\leq V(x)$ when $x$ is in a small neighborhood of $x^*$.  However, when $x\rightarrow x^*$, the linear term $\frac{1}{2} \left[V'(x^{*}-) + V'(x^{*}+)  \right](x-x^*)$ will dominate the quadratic term $\alpha(x-x^*)^2$. Therefore, when $x$ and $x^*$ are close enough, we could choose sufficiently large $\alpha$ such that $\psi(x)\leq V(x)$. It is now clear that $\alpha$ can take any value. \\
Now, apply the viscosity property at $x^*$, we have
\beq
F\left(x^*, V(x^*), \frac{1}{2} \left[V'(x^{*}-) + V'(x^{*}+)  \right], 2 \alpha\right) \ge 0,
\eeq
which is impossible by the free choice of the parameter $\alpha$.  \hfill$\Box$

Recall the Legendre-Fenchel transformation of the function $f(c,x)$ is defined as
\beq
p(x, \zeta) = \max_{c \ge 0} \left\{ f(c,x) - c \zeta \right\}.
\eeq

\begin{lem}
\label{lem:p}
The function $p(x, \zeta)$ is a convex function of $\zeta$.
\end{lem}

\proof
It is clear that
$p(x,\zeta)$ is continuous on $x$.  Recall the definition of $I(x,\zeta)$,  we have ${\partial f\over \partial c}(I(x,\zeta),x)=\zeta$. By ${\partial ^2f \over \partial c^2}(c,x) < 0$, we have ${\partial I \over \partial \zeta}(x,\zeta) < 0$. Rewrite $p(x,\zeta) = f(I(x, \zeta), x) - I(x,\zeta) \zeta$. Then,
\begin{eqnarray*}
{\partial p \over \partial \zeta}(x,\zeta) &=& {\partial f \over \partial c}(I(x,\zeta), x) {\partial I \over \partial \zeta}(x, \zeta) - {\partial I \over \partial \zeta}(x, \zeta) \zeta - I(x,\zeta) \\
& = & - I(x,\zeta),
\end{eqnarray*}
and ${\partial^2 p \over \partial \zeta^2}(x,\zeta) = -  {\partial I \over \partial \zeta}(x, \zeta) > 0$.  \hfill$\Box$

By Proposition \ref{prop:Case1-vis} and the structure of equation (\ref{eq:Case1-vis}), we define the unconstrained domain ${\cal U}$ be the set of $x$ such that $V(x)$ is the viscosity solution of 
\be
\label{eq:Case1- ODE no constraint-appen}
\delta {V}(x) = -{(\mu-r)^2\over 2\sigma^2}{(V')^2(x)\over V''(x)}+ p(x, V'(x))+ rx V'(x). 
\ee
Also, define the constrained domain ${\cal B}$ be set of $x$ such that $V(x)$ is the viscosity solution of 
\begin{equation}
\label{eq:Case1-ODE with constraint-appen}
\delta {V}(x) =   (\mu-r) g(x) V'(x) + \frac{1}{2} \sigma^2 g^2(x) {V}''(x) +p(x, V'(x))+ rx V'(x).    
\end{equation}
Note that $(0,\infty)$ will be divided into three parts: (1) The unconstrained domain ${\cal U}$; (2) the constrained domain ${\cal B}$; and (3) the connection points in $cl({\cal U}) \bigcap cl({\cal B})$.

{\bf Proof of  Theorem \ref{thm:Case1-smooth}}: 

We divide the proof into several steps.

{\bf Step 1:} By Proposition \ref{prop:Case1-vis},  the value function $V(x)$ is strictly increasing, strictly concave function.  Moreover,  $V(x)$ is the unique viscosity solution of (\ref{eq:Case1-vis}).

{\bf Step 2:} We show $V(x)$ is $C^2$ at $x\in \cal B$.

For any $x\in \cal B$,  $V(x)$ is the viscosity solution of 
\beq
\delta {V}(x) =   (\mu-r) g(x) V'(x) + \frac{1}{2} \sigma^2 g^2(x) {V}''(x) + p(x, V'(x)) + rx V'(x).   
\eeq
Recall $g(x)\geq L, \forall x>0$ .  Then $\frac{1}{2} \sigma^2 g^2(x) $,  the coefficients of $V''(x)$ satisfies
\beq
\frac{1}{2} \sigma^2 g^2(x)\geq \frac{1}{2} \sigma^2 L^2>0, \quad \forall x\in \cal B
\eeq
which is uniformly positive. Hence,  when $x\in \cal B$,  $V(x)$ is $C^2$ due to the the equation is clearly non-degenerate,  see Krylov (1987).

{\bf Step 3:}  We show that the value function $V(x)$ is $C^1$ at $x\in {\cal U}$.

 Since $V$ is increasing and concave, we define its right and left derivative:
\begin{equation}
V'(x\pm)=\lim_{h\rightarrow 0+} {V(x\pm h)-V(x)\over \pm h} \geq 0
\end{equation}
for all $x>0$.  Note that $0\leq V_x(x+)\leq V_x(x-)<\infty$ for all $x>0$.  By Lemma \ref{lem: c1-property} ,  in order to show that $V(x)$ is $C^1$,  it suffices to show that $V(x)$ is differentiable, i.e. $V'(x-)=V'(x+)$.

Now we prove the result by contradiction. Assume $V'(x_0+)<V'(x_0-)$ for some $x_0 \in {\cal U}$ .  Set $\eta$ satisfies $V'(x_0+)<\eta<V'(x_0-)$.  Define
\begin{equation}
\label{eq: def of phi}
\phi(x)=V(x_0)+\eta(x-x_0)-m(x-x_0)^2
\end{equation}
for $m>0$.  It is clear that $\phi(x_0)=V(x_0)$, $\phi'(x_0)=\eta$ and $\phi''(x_0)=-2m$.

By concavity of $V(x)$, when $0<x_0-x<{1\over m}[V'(x_0-)-\eta]$, we have
\beq
V(x)&\leq & V(x_0)+V'(x_0-)(x-x_0) \nonumber
\eeq
Plugging (\ref{eq: def of phi}) into it, we get
\begin{eqnarray}
\label{eq:test-left-X}
\nonumber
V(x) & \leq &\phi(x)+(V'(x_0-)-\eta)(x-x_0)+m(x-x_0)^2\nonumber\\
&<& \phi(x)
\end{eqnarray}

Similarly,  when $0<x-x_0<{1\over m} [\eta-V'(x_0+)]$, we have
\begin{eqnarray}
\label{eq:test-right-X}
\nonumber
V(x)&\leq & V(x_0)+V'(x_0+)(x-x_0)\\ \nonumber
&=&\phi(x)+(V'(x_0+)-\eta)(x-x_0)+m(x-x_0)^2\\
&<& \phi(x)
\end{eqnarray}

Basically,  (\ref{eq:test-left-X}) and (\ref{eq:test-right-X}) imply that $V(x)<\phi(x)$ in a small neighborhood of $x_0$, therefore we may use $\phi(x)$ as the desired test function at $x=x_0$.  Since when $x\in {\cal U}$, $V(\cdot)$ is a viscosity solution of (\ref{eq:Case1- ODE no constraint-appen}),  using the definition of viscosity subsolution at $x_0$, we have
\begin{eqnarray}
\nonumber
0&\geq & \delta \phi(x_0)+{(\mu-r)^2\over 2\sigma^2}{(\phi'(x_0))^2\over \phi''(x_0)}- p(x_0, \phi'(x_0)) -rx_0  \phi'(x_0)  \\\nonumber
&=&\delta V(x_0)-{\mu^2\eta^2\over 4\sigma^2m}- p(x_0, \eta) - rx_0 \eta.
\end{eqnarray}
Sending $m\rightarrow\infty$,  we have 
\be
\label{eq: g1-X}
0 \geq \delta V(x_0)-p(x_0, \eta) - rx_0 \eta.
\ee
for $\forall \eta\in (V'(x_0+),V'(x_0-))$

On the other hand, since $V(\cdot)$ is concave,  it is second differentiable almost everywhere. Then there exists $\{x_n\}_{\{n\geq 0\}}$ increases to $x_0$ such that $V(\cdot)$ is $C^2$ at all $x_n$. Then by (\ref{eq:Case1- ODE no constraint-appen}), we have
\begin{eqnarray}
\nonumber
0&= &\delta {V}(x_n)+{(\mu-r)^2\over 2\sigma^2}{V'(x_n)^2\over V''(x_n)}- p(x_n, V'(x_n))- rx_n V'(x_n)   \\
&\leq & \delta V(x_n)- p(x_n, V'(x_n)) - rx_n V'(x_n)
\end{eqnarray}
Sending $x_n \uparrow x_0$, we get
\begin{eqnarray}
\label{eq:g2-X}
0\leq \delta V(x_0)- p(x_0, V'(x_{0}-) - rx_0 V'(x_0-)
\end{eqnarray}
Similarly, by choosing $x_n \downarrow x_0$, we get
\begin{eqnarray}
\label{eq:g3-X}
0\leq \delta V(x_0)- p(x_0, V'(x_{0}+) - rx_0 V'(x_{0}+).
\end{eqnarray}
Define $q(x_0, \eta)= p(x_0, \eta) + rx_0\eta$. It is clear that ${\partial^2 q \over \partial \eta^2}(x_0, \eta) = {\partial^2 p \over \partial \eta^2}(x_0, \eta) > 0$ by Lemma \ref{lem:p}.   Then $q(x_0, \eta)$  is a convex function of $\eta$.  By (\ref{eq: g1-X}),  we have
\be
\label{eq: g geq}
q(x_0,\eta) \geq \delta V(x_0),  \quad \forall \eta\in (V'(x_0+),V'(x_0-))
\ee
On the other hand, by convexity of $q(x_0, \eta)$ on $\eta$, we have
\beq
q(x_0, \eta) \leq \min (q(x_0, V'(x_0-)), q(x_0, V'(x_0+))),  \quad \forall \eta\in (V'(x_0+),V'(x_0-))
\eeq
Use (\ref{eq:g2-X}) and (\ref{eq:g3-X}), we have
\be
\label{eq: g leq}
q(x_0, \eta) \leq \delta V(x_0),  \quad \forall \eta\in (V'(x_0+),V'(x_0-)).
\ee
Now,  combining (\ref{eq: g geq}), (\ref{eq: g leq}),  we conclude that $h(\eta)$ is a constant on $\eta\in  (V'(x_0+),V'(x_0-))$. However,   since 
\begin{equation}
{\partial^2 q \over \partial \eta^2}(x_0, \eta)  > 0,
\end{equation}
Hence, $q(x_0, \eta)$ can not  be a constant function on $\eta\in  (V'(x_0+),V'(x_0-))$. We therefore conclude that the value function $V$ is $C^1$  when $x\in \cal U$.

{\bf Step 4:} We show the value function $V$ is $C^2$ at $x\in \cal U$.

For this purpose, we define the dual transformation:
\begin{equation}
\label{eq: def-dual-X}
v(y):= \max_{x>0}(V(x)-xy),  \quad \lim\limits_{x\rightarrow +\infty} V'(x)<y< V'(0)
\end{equation}
Then $v(\cdot)$ is a decreasing convex function on $(\lim\limits_{x\rightarrow +\infty} V'(x),V'(0))$. Since $V'(\cdot)$ is strictly decreasing, we denote the inverse function of $V'(x)=y$ by $
I(y)=x$. Then $I(\cdot)$ is decreasing and mapping $(\lim\limits_{x\rightarrow +\infty} V'(x),V'(0))$ to $(0,\infty)$. Also, from (\ref{eq: def-dual-X}), we get
\begin{equation}
\label{eq:dual-plug-X}
v(y)=[V(x)-xV'(x)]|_{x=I(y)}=V(I(y))-yI(y)
\end{equation}
Differentiate (\ref{eq:dual-plug-X}) once and twice,  we get
\begin{equation}
\label{eq:dual-first-X}
v'(y)=V'(I(y))I'(y)-yI'(y)-I(y)=-I(y)
\end{equation}
and
\begin{equation}
\label{eq:dual-second-X}
v''(y)=-I'(y)=-{1\over V''(I(y))}
\end{equation}
Combining (\ref{eq:dual-plug-X}) and (\ref{eq:dual-first-X}), we get
\begin{equation}
\label{eq:V-dual-X}
V(I(y))=v(y)-yv(y)'
\end{equation}
Now,  plugging (\ref{eq:dual-first-X}),(\ref{eq:dual-second-X}),(\ref{eq:V-dual-X} ) into (\ref{eq:Case1- ODE no constraint-appen}), we get
\begin{equation}
\label{eq:transODE-X}
\delta(v(y)-yv'(y))={(\mu-r)^2\over 2\sigma^2}y^2v''(y)+ p(-v'(y),y) - r v'(y) y,  \quad  \lim\limits_{x\rightarrow +\infty} V'(x)<y<V'(0)
\end{equation}

Equation (\ref{eq:transODE-X}) is a quasilinear ODE, which only degenerates at $y=0$. Since $V'(x) > 0$ for all $x > 0$, we have $0 \notin (\lim\limits_{x\rightarrow +\infty} V'(x), V'(0))$. Then, the coefficient of $v''(y)$ in the above equation is $\frac{(\mu-r)^2}{2 \sigma^2} y^2$, which is nonzero. It follows that
\begin{equation}
v\in C^2(\lim\limits_{x\rightarrow +\infty} V'(x),V'(0)).
\end{equation}
By (\ref{eq:dual-second-X}),  we get $V(\cdot)$ is $C^2$ when $x\in \cal U$.

{\bf Step 5:} In Step 2 and Step 4,  We have proved that the value function $V$ is $C^2$ when $x\in \cal U$ and $x\in \cal B$.  In this step, we show that $V$ is $C^2$ when $x$ is  the  connection point in $cl({\cal U}) \bigcap cl({\cal B})$.

Assume $x^*$ is  the  connection point in $cl({\cal U}) \bigcap cl({\cal B})$.  Our goal is to show that $V$ is $C^2$ at $x^*$. Without loss of generality,  we assume that the left neighborhood of $x^*$ is $\cal U$ and right neighborhood of $x^*$  is $\cal B$.  That is, we have 
\begin{equation}
\label{eq:ODE-left}
\delta {V}(x^*-) = -{(\mu-r)^2\over 2\sigma^2}{V'(x^*-)^2\over V''(x^*-)}+ p(x^*-, V'(x^*-)) + rx^*V'(x^*-)
\end{equation}
and 
\begin{eqnarray}
\label{eq:ODE-right}
\delta {V}(x^*+) &=&  (\mu-r) g(x^*) V'(x^*+) + \frac{1}{2} \sigma^2 (g(x^*))^2 {V''}(x^*+)\\\nonumber
 &&+ p(x^*+, V'(x^*+)) + rx^*V'(x^*+).
\end{eqnarray}
Since $V$ is continuous on $(0,\infty)$, we have 
\begin{equation}
\label{eq:value-match}
V(x^*+)=V(x^*-).
\end{equation}
Moreover,  since $V$ is the viscosity solution of (\ref{eq:Case1-vis}) and $V$ is $C^2$  when $x\in \cal U$ and $x\in \cal B$,  apply Lemma \ref{lem:smooth-fit}, we obtain
\begin{equation}
\label{eq: der-match}
V'(x^*+)=V'(x^*-).
\end{equation}
Next,  we use (\ref{eq:value-match}) (\ref{eq: der-match}) and $p$ is a continuous function,  compare the terms in (\ref{eq:ODE-left})and (\ref{eq:ODE-right}),  we have
\be
\label{eq: remaining terms }
-{(\mu-r)^2\over 2\sigma^2}{V'(x^*-)^2\over V''(x^*-)}=(\mu-r) g(x^*) V'(x^*+) + \frac{1}{2} \sigma^2 (g(x^*))^2 {V''}(x^*+)
\ee

Moreover,  when $x\in \cal U$,  for the optimal $\pi^*$, we have
\begin{equation}
\label{eq:opcontrol-unconstrained}
 \pi^*(x^*-)=-{(\mu-r)V'(x^*-)\over \sigma^2V''(x^*-)}=g(x^*)
\end{equation}
Plugging (\ref{eq:opcontrol-unconstrained}) into (\ref{eq: remaining terms }),  we get
\begin{equation}
{\mu-r\over 2}g(x^*)V'(x^*-)=({\mu-r})g(x^*)V'(x^*+)+ \frac{1}{2} \sigma^2 (g(x^*))^2 {V}''(x^*+)\nonumber
\end{equation}
Use (\ref{eq: der-match}), we get
\begin{equation}
\label{eq: seder-match}
-{\mu-r\over 2}g(x^*)V'(x^*-)= \frac{1}{2} \sigma^2 (g(x^*))^2 {V}''(x^*+)
\end{equation}
Finally,  combine (\ref{eq:opcontrol-unconstrained}) and (\ref{eq: seder-match}), we get
\begin{equation}
V''(x^*+)=V''(x^*-)
\end{equation}
Since $x^*$ is arbitrary connection point in $cl({\cal U}) \bigcap cl({\cal B})$,  we have shown that $V(x)$ is $C^2$ on all  the  connection points in $cl({\cal U}) \bigcap cl({\cal B})$.

{\bf Step 6:} (Verification) In this step, we show that if $\overline{V}(x)$ is the concave $C^2$ solution of (\ref{eq:Case1-HJB}),  then we must have $\overline{V}(x)=V(x)$ . Fix $T>0$, for arbitrary $(c_t, \pi_t) \in {\cal A}(x)$, define
\beq
\tau_n=(T-{1\over n})^+\wedge \inf\{s\in [0,T]; X_s\geq n\ or\ X_s\leq {1\over n}\ or\ \int_0^s \pi^2_u  du=n\}
\eeq
apply Ito's formula to $e^{-\delta t}\overline{V}(X_t)$ on $[0,T\wedge \tau_n]$, we have:
\beq
E e^{-\delta (T\wedge \tau_n)}\overline{V}(X_{T\wedge \tau_n})=&&\overline{V}(x)+E\int_0^{T\wedge \tau_n} e^{-\delta t}[\overline{V}'(X_t)(\pi_t (\mu-r)+rX_t-C_t)\\
&&+{1\over 2}\overline{V}''(x)\sigma^2\pi_t^2-\delta \overline{V}(X_t)]dt
\eeq
Since $\overline{V}$ is the $C^2$ solution of (\ref{eq:Case1-HJB}), we have
\beq
E e^{-\delta (T\wedge \tau_n)}\overline{V}(X_{T\wedge \tau_n})\leq \overline{V}(x)-E\int_0^{T\wedge \tau_n} e^{-\delta t}f(c_t,X_t)dt
\eeq
First send $n\to \infty$, then send $T\to \infty$, since $\overline{V}$ is concave, it is easy to see the transversality condition $\lim_{T\to \infty} Ee^{-\delta T} \overline{V}(X_T)=0$.  Therefore, we have
\beq
E\int_0^\infty e^{-\delta t}f(c_t,X_t)dt \leq \overline{V}(x)
\eeq
since $(c_t,\pi_t)$ is arbitrary, we have $\overline{V}(x)\geq V(x)$.

On the other hand, if we choose $(c_t^*,\pi_t^*) \in {\cal A}(x)$ such that
\beq
c_t^*=I(X_t^*,\overline{V}'(X_t^*)), \pi^*_t=\min(g(X^*_t),{-(\mu-r)\overline{V}'(X_t^*)\over \sigma^2\overline{V}''(X_t^*)})
\eeq
where $I(x,\zeta)$ satisfies ${\partial f\over \partial c}(I(x,\zeta),x) = \zeta$. Then repeat the above process, we get
\beq
 \overline{V}(x)=E\int_0^\infty e^{-\delta t}f(c^*_t,X^*_t)dt \leq V(x)
\eeq
The feedback control part is standard. We therefore conclude the proof.
$\hfill \Box$

{\bf Proof of Lemma \ref{lem:proportion}.}

It suffices to show that ${g(x)\over x}$ is a decreasing function when $x> 0$. Note that 
\beq
({g(x)\over x})'={xg'(x)-g(x)\over x^2}
\eeq
We will show that $h(x):=xg'(x)-g(x)\leq 0$ and therefore complete the proof. Note that $\lim_{x\rightarrow 0}h(x)=\lim_{x\rightarrow 0} xg'(x)-g(0)$. Since $g(x)$ is a Lipschitz function, therefore, $g'(x)\leq K, \forall x>0$ for some positive constant $K$. Therefore, $\lim_{x\rightarrow 0} xg'(x)=0$. We then have $\lim_{x\rightarrow 0}h(x)=-g(0)\leq 0$. Moreover, since $g(x)$ is a concave function,
\beq
h'(x)=xg''(x)\leq 0.
\eeq
Therefore, $h(x)$ is a decreasing function. We conclude $h(x)\leq \lim_{x\rightarrow 0}h(x)\leq 0, \forall x> 0$. $\hfill \Box$

We first show the following result in order to prove Lemma \ref{lem:unconstrained}.

\begin{lem}
\label{lem:V'(x)}
$\lim\limits_{x\rightarrow 0} V'(x)=+\infty$.
\end{lem}

\proof It is clear that for $x>0$, $(c_t,\pi_t)=(rX_t,0)\in A(x)$. Therefore
\beq
V(x)-V(0)&\geq& E\int_0^{\infty} e^{-\delta t}[f(rX_t,X_t)-f(0,0)]dt\\
         &=& E\int_0^{\infty} e^{-\delta t}[f(rx,x)-f(0,0)]dt
\eeq
Since in Theorem \ref{thm:Case1-smooth}, it is already shown that $V(x)$ is $C^2$ smooth. Then by Fatou's lemma and the Inada's condition on $f$, we get
\beq
\lim_{x\rightarrow 0} V'(x)&=&\lim_{x\rightarrow 0} {V(x)-V(0)\over x}\\
&\geq & E\int_0^{\infty} e^{-\delta t} \liminf_{x\rightarrow 0}[{f(rx,x)-f(0,0)\over x}]dt\\
&=&+\infty
\eeq
$\hfill \Box$

{\bf Proof of Lemma \ref{lem:unconstrained}.}

 Assume not, then there exists a sequence $x_n \rightarrow 0$ such that $\pi^*(x_n)=g(x_n)$ satisfies
\be
\label{eq:ODE constrained near 0}
\delta {V}(x_n) =   (\mu-r) g(x_n) V'(x_n) + \frac{1}{2} \sigma^2 g^2(x_n) {V}''(x_n) +p(x_n, V'(x_n))+ rx_n V'(x_n).  
\ee
By the definition of $\cal{B}$, we have $\frac{\mu-r}{\sigma^2} \left( - \frac{V'(x_n)}{V''(x_n)} \right) > g(x_n)$, therefore (\ref{eq:ODE constrained near 0}), the nonnegativity of $p(x,\zeta)$ and $V'(x)$ implies that
\beq
\delta {V}(x_n) &\geq& {1\over 2}(\mu-r) g(x_n) V'(x_n)+p(x_n, V'(x_n))+ rx_n V'(x_n)\\
&\geq & {1\over 2}(\mu-r) L V'(x_n).
\eeq
Sending $n\rightarrow \infty$, By the continuity of $V$, $\delta {V}(x_n)$ converges to $V(0)=0$. However, by Lemma \ref{lem:V'(x)}, ${1\over 2}(\mu-r) L V'(x_n)
\rightarrow +\infty$,  contradiction.
$\hfill \Box$
{\bf Proof of Theorem \ref{thm: Two region}:}

When $g(x)\equiv L$, the ordinary differential equation for $V(x)$ in the unconstrained and constrained region are
\be
\label{eq:V-unconstrained}
\delta V(x) = \theta \frac{ (V'(x)^2}{-V''(x)}+ p(x, V'(x))+ rxV'(x) , \theta \equiv \frac{(\mu-r)^2}{2\sigma^2}.
\ee 
and 
\be
\label{eq:V-constrained}
\delta {V}(x) =   (\mu-r) L V'(x) + \frac{1}{2} \sigma^2 L^2 {V}''(x) +p(x, V'(x))+ rx V'(x). 
\ee
We define a function
\be
Y(x) = (\mu -r)V'(x) + \sigma^2 L V''(x), x > 0.
\ee
Then, $Y(x) > 0, \forall x \in {\cal U}$, and $Y(x) < 0$ for any $x \in {\cal B}$.

{\bf Step 1:} In the unconstrained region, the value function $V(\cdot )$ satisfies the ODE \ref{eq:V-unconstrained}, differentiate this ODE once and twice, we get
\beq
\delta V'=-2\theta V'+{\theta (V')^2V'''\over (V'')^2}+{\partial p\over \partial x}(x,V')+{\partial p\over \partial \zeta}(x,V')V''+rV'+rxV''
\eeq
and 
\beq
\delta V''&=&-2\theta V''+{\theta (V')^2V''''\over (V'')^2}+{2\theta V'V'''\over (V'')^3}[(V'')^2-V'V''']+{\partial^2 p\over \partial x^2}(x,V')+{\partial^2 p\over \partial x \partial \zeta}(x,V')V''\\
&&+[{\partial^2 p\over \partial x \partial \zeta}(x,V')+{\partial^2 p\over \partial \zeta^2}(x,V')V'']V''+{\partial p \over \partial \zeta}(x,V')V'''+2rV''+rxV'''.
\eeq
By the definition of $Y(x)$, the last two equations imply that
\beq
\delta Y&=&-2\theta Y+{\theta (V')^2\over (V'')^2}Y''+{2\theta V'V'''\over (V'')^3}[{V''\over \sigma^2 L}Y-{V'\over \sigma^2L}Y']+[{\partial p \over \partial \zeta}(x,V')+rx]Y'+rY\\
&&+(\mu-r){\partial p \over \partial x}(x,V')+\sigma^2L[{\partial^2 p\over \partial x^2}(x,V')+2{\partial^2 p\over \partial x \partial \zeta}(x,V')+{\partial^2 p\over \partial \zeta^2}(x,V')(V'')^2+rV'']\\
&=&-2\theta Y+{\theta (V')^2\over (V'')^2}Y''+{2\theta V'V'''\over (V'')^3}[{V''\over \sigma^2 L}Y-{V'\over \sigma^2L}Y']+[{\partial p \over \partial \zeta}(x,V')+rx]Y'+rY\\
&&+(\mu-r){\partial p \over \partial x}(x,V')+\sigma^2L[{\partial^2 p\over \partial x^2}(x,V')+2{\partial^2 p\over \partial x \partial \zeta}(x,V')\\
&&+{\partial^2 p\over \partial \zeta^2}(x,V')({1\over \sigma^2 L}(Y-(\mu-r) V'))^2+{r\over \sigma^2L}(Y-(\mu-r)V')]\\
&=& -2\theta Y+{\theta (V')^2\over (V'')^2}Y''+{2\theta V'V'''\over (V'')^3}[{V''\over \sigma^2 L}Y-{V'\over \sigma^2L}Y']+[{\partial p\over \partial \zeta}(x,V')+rx]Y'+2rY\\
&&+{1\over \sigma^2L}{\partial^2 p\over \partial \zeta^2}(x,V')[Y^2-2(\mu-r)V'Y]+(\mu-r){\partial p\over \partial x}(x,V')\\
&&+\sigma^2L[{\partial^2 p \over \partial x^2}(x,V')+2{\partial^2 p\over \partial x \partial \zeta}(x,V')]+{1\over \sigma^2L}{\partial^2 p\over \partial \zeta^2}(x,V')(\mu-r)^2(V')^2-r(\mu-r)V'
\eeq
We then define an elliptic operator on the unconstrained region by
\beq
{\cal L}^{\cal U}[y]&\equiv& -{\theta (V')^2\over (V'')^2}y''-{2\theta V'V'''\over (V'')^3}[{V''\over \sigma^2 L}y-{V'\over \sigma^2L}y']-[{\partial p\over \partial \zeta}(x,V')+rx]y'+(2\theta+\delta-2r) y\\
&&-{1\over \sigma^2L}{\partial^2 p\over \partial \zeta^2}(x,V')(y^2-2(\mu-r)V'y)-(\mu-r){\partial p\over \partial x}(x,V')\\
&&-\sigma^2L[{\partial^2 p\over \partial x^2}(x,V')+2{\partial^2 p\over \partial x\partial \zeta}(x,V')]-{1\over \sigma^2L}{\partial^2 p \over \partial \zeta^2}(x,V')(\mu-r)^2(V')^2+r(\mu-r)V'
\eeq
Therefore, ${\cal L}^{\cal U}[Y]=0$ in ${\cal U}$.

{\bf Step 2}. In the constrained region ${\cal B}$, by differentiate the ODE (\ref{eq:V-constrained}) once and twice, we get
\beq
\delta {V}'=   (\mu-r) L V'' + \frac{1}{2} \sigma^2 L^2 {V}''' +{\partial p\over \partial x}(x,V')+{\partial p\over \partial \zeta}(x,V')V''+rV'+rxV''.
\eeq
and 
\beq
\delta {V}'' &=&   (\mu-r) L V''' + \frac{1}{2} \sigma^2 L^2 {V}''''+{\partial^2 p\over \partial x^2}(x,V')+{\partial^2 p\over \partial x \partial \zeta}(x,V')V''\\
&&+[{\partial^2 \over \partial x\partial \zeta}(x,V')+{\partial^2 p\over \partial \zeta^2}(x,V')V'']V''+{\partial p\over \partial \zeta}(x,V')V'''+2rV''+rxV'''
\eeq
Again, by the definition of $Y(x)$, we have
\beq
\delta Y&=&(\mu-r)LY'+{1\over 2\sigma^2 L^2}Y''+[{\partial p\over \partial \zeta}(x,V')+rx]Y'+2rY\\
&&+{1\over \sigma^2L}{\partial^2 p\over \partial \zeta^2}(x,V')[Y^2-2(\mu-r)V'Y]+(\mu-r){\partial p\over \partial x}(x,V')+\sigma^2L[{\partial^2 p \over \partial x^2}(x,V')\\
&&+2{\partial^2 p\over \partial x\partial \zeta}(x,V')]+{1\over \sigma^2L}{\partial^2p \over \partial \zeta^2}(x,V')(\mu-r)^2(V')^2-r(\mu-r)V'
\eeq
Similarly, we define an elliptic operator
\beq
{\cal L}^{\cal B}[y]&=&-{1\over 2\sigma^2 L^2}y''-(\mu-r)Ly'-[{\partial p\over \partial \zeta}(x,V')+rx]y'-2ry-{1\over \sigma^2L}{\partial^2 p \over \partial \zeta^2}(x,V')[y^2\\
&&-2y(\mu-r)V']-(\mu-r){\partial p\over \partial x}(x,V')-\sigma^2L[{\partial^2 p\over \partial x^2}(x,V')+2 {\partial^2 \over \partial x\partial \zeta}(x,V')]\\
&&-{1\over \sigma^2L}{\partial^2 p\over \partial \zeta^2}(x,V')(\mu-r)^2(V')^2+r(\mu-r)V'
\eeq
Then ${\cal L}^{\cal B}[Y]=0$ in ${\cal B}$.

{\bf Step 3}. By computation, we get 
\beq
{\cal L}^{\cal B}[0]={\cal L}^{\cal U}[0]&=&-(\mu-r){\partial p\over \partial x}(x,V')-\sigma^2L[{\partial^2 p\over\partial x^2}(x,V')+2{\partial ^2 \over \partial x\partial \zeta}(x,V')]\\
&&-{1\over \sigma^2L}{\partial ^2 p\over \partial \zeta^2}(x,V')(\mu-r)^2(V')^2+r(\mu-r)V'
\eeq
Note that this function is exactly $m(x)$ defined in the theorem.

{\bf Step 4.}  By Lemma \ref{lem:unconstrained}, there exists a real number $x_1 > 0$ such that $(0, x_1) \subseteq {\cal U}$ and $Y(x_1) = 0$. 
Since ${\cal U} \ne (0, \infty)$, we have $x_1<\infty$. We show that $(x_1, \infty) \subseteq {\cal B}$ by a contradiction argument. 

Assume that, there exists $x_2 > x_1$ such that $(x_1, x_2) \subseteq {\cal B}$ and $Y(x_2) = 0$. Moreover, there exists $x_3 > x_2$ such that $(x_2, x_3) \subseteq {\cal U}$. We show this is impossible and thus finish the proof.

We first show that the constant function $y=0$ is {\em not} the supersolution for ${\cal L}^{{\cal B}}[y ]= 0$ in the region $(x_1, x_2)$. The reason is as follows. Otherwise, since ${\cal L}^{{\cal B}}[Y] = 0$ in the region $(x_1, x_2) \subseteq {\cal B}$ and $Y(x_1) = Y(x_2) = 0$, then by the comparison principle, $Y(x) \le y=0$ for $x \in (x_1, x_2)$. However, by its definition of ${\cal B}, Y(x) > 0$ for all $x \in (x_1, x_2)$. This contradiction show that the constant funciton $y=0$ is not a supersolution of ${\cal L}^{{\cal B}}[y ]= 0$ in the region $(W_1, W_2)$.
Therefore, there exists some $x_0 \in (x_1, x_2)$ such that, at $x = x_0$,
\beq
{\cal L}^{{\cal B}}[0 ] = m(x_0) < 0.
\eeq

We divide the proof into two situations because of the single crossing condition 

{\bf Case 1.}  The function $m(x)$ does not change sign at all in $(0, \infty)$.

In this case, $m(x) < 0$ for all $(0,\infty)$. We now consider the region $(x_2, x_3) \in {\cal U}$ and the operator ${\cal L}^{{\cal U}}$. Since ${\cal L}^{{\cal U}}[0] \le 0$ in this small region, the constant function $y=0$ is the subsolution for ${\cal L}^{{\cal U}}[0] = 0$. Since  $Y(x_2) = Y(x_3) = 0$, by the comparison principle, we obtain $Y(x) \ge 0, \forall x \in (x_2, x_3)$, which is impossible since $Y(x)$ is strictly negative over the region $(x_2, x_3) \subseteq {\cal U}$, the unconstrained region. 

{\bf Case 2.}  The function $m(x)$ change the sign once and only once from positive to negative on $(0,\infty)$.

By $m(x_0)<0$ and the condition on $m(x)$, the function $m(x)$ must be negative for all $x > x_0$. In particular, $m(x) < 0, \forall x \in (x_2, x_3)$. Following the same proof as in Case 1, the constant $y=0$ is the subsolution for ${\cal L}^{{\cal U}}[0] = 0$. It implies that $Y(x) \ge 0, \forall x \in (x_2, x_3)$. This leads a contradiction again by the definition of ${\cal U}$.

By the above proof, we have shown that $(x_1, \infty) = {\cal B}$ by a contradiction argument.  
$\hfill \Box$

{\bf Proof of Proposition \ref{ex: utility only c}:}

 We check that this utility function satisfies the condition in Theorem \ref{thm: Two region}. In this case,
\beq
p(x,\zeta)=u(K (\zeta))-K(\zeta)\zeta
\eeq
where $K(\zeta)=(u')^{-1}(\zeta)$. Clearly, $p(x,\zeta)$ is independent of $x$, hence ${\partial p \over \partial x}(x,\zeta)={\partial^2 p \over \partial x^2}(x,\zeta)={\partial^2 p \over \partial x \partial \zeta}(x,\zeta)=0$. Moreover, by Lemma \ref{lem:p}, ${\partial^2 p \over \partial \zeta^2}(x,\zeta)=-K'(\zeta)>0$. Therefore, 
\be
m(x)=-{1\over \sigma^2L}{\partial^2 p \over \partial \zeta^2}(x,V'(x))\mu^2(V'(x)))^2<0
\ee
satisfies the first condition of $m(x)$ in Theorem \ref{thm: Two region}. Therefore, the two-regions property holds. 
$\hfill \Box$

{\bf Proof of Proposition \ref{co:2}:}

We verify that this utility function satisfies the condition in Theorem \ref{thm: Two region}.
 In this case $p(x,\zeta)=u(x).$ 
It is clear that ${\partial p \over \partial \zeta}(x,\zeta)={\partial^2 p \over \partial x\partial \zeta}(x,\zeta)={\partial^2 p \over \partial \zeta^2}(x,\zeta)=0$. Therefore, 
\beq
m(x)=-\mu u'(x)-\sigma^2Lu''(x)
\eeq
The proposition follows from Theorem \ref{thm: Two region}. \hfill$\Box$

{\bf Proof of Proposition \ref{prop:additive}:}

It follows from Theorem \ref{thm: Two region} and a direct calculation.  \hfill$\Box$

\newpage

\section*{Declarations}
{\bf Funding and/or Conflicts of interests/Competing interests}: The authors have no conflicts of interest to declare that are relevant to the content of this article.

\end{document}